\documentstyle[aps,manuscript]{revtex}

\draft

\begin{document}

\title{Morphological instabilities of a thin film on a Penrose lattice:
a Monte Carlo study 
}

\author{ N. Olivi-Tran, A.Boulle, A.Gaudon and A.Dauger 
}
\address{SPCTS, UMR-CNRS 6638, ENSCI, 47 avenue Albert Thomas, 87065 Limoges cedex,
France}

\maketitle
\date{today}

\begin{abstract}
We computed by a Monte Carlo method the thermal relaxation of a polycrystalline thin film
deposited on a Penrose  lattice.
The thin film was modelled by a 2 dimensional array of elementary
domains, which have each a given height. During the Monte Carlo
process, the height of each of these elementary domains
is allowed to change as well as their crystallographic orientation.
After equilibrium is reached at a given numerical temperature,
all elementary domains have changed their orientation into
the same one and small islands appear, preferentially on the
domains of the Penrose lattice located in the center of heptagons.
This method is a new numerical approach to study the influence
of the substrate and its defects on the islanding process of polycrystalline films.
 
\end{abstract}
\vfill
\pacs{Keywords: Islanding, thin film, numerical simulation, geometrical frustration}

\section{Introduction}
In recent years, the formation of mesoscopic 
structures on crystal surfaces has become a subject of intense experimental
and theoretical study. 
Generally, for non periodically ordered nanostructures, the increasing specific 
area is favorable in order to enhance the physical properties (in optics,
semiconducting etc) owing to the increased number of active sites \cite{shalaev}.

We will study here the evolution of thin film deposited on a substrate with non periodical
defects.
For thin films deposited on a crystalline substrate with defects,
the epitaxial growth of such thin films is always accompanied
by the formation of elastic strain emanating from the film-substrate
interface.
The knowledge and the control of defects and their associated strain fields
is therefore a key issue for the fabrication of thin film
based devices.

Several  methods (e.g. sol gel processing \cite{brinker}) allow one to obtain, after a 
first heat treatment (stage I), thin films of nanometric thickness, 
made of a large amount nanocrystals of random orientation.
At this stage, the thin film thickness is much larger than the mean size of 
these nanocrystals.
After a second heat treatment at higher temperature(stage II), thermal annealing 
induces grain growth. At this stage, the size of the crystals 
reaches the thickness of the thin film. Moreover, the crystals with the lowest
interfacial energy, with respect to the substrate, are expected to grow
to the detriment of crystals with higher energies \cite{thompson}.
Simultaneously, the film is submitted to fragmentation into
more or less interconnected islands in order to reduce the total energy
and hence to reach a more stable state \cite{miller}.

The numerical model deals with a polycrystalline thin film
deposited on a substrate with defects represented by the vertices
of a Penrose tiling .
 The computation
starts at stage II of the thermal relaxation of the experimental
thin film, i.e. the mono-crystals composing the thin film have the thickness of the 
thin film.
The aim of this model is to analyze the influence of the defects of the substrate on
 the resulting islands after  fragmentation of the thin film occurred.

In section II, we present the numerical procedure. And in section
III, results are discussed. Finally, we conclude in section IV.
 
 \section{Numerical procedure}
We modelled a thin polycrystalline film deposited on a
substrate with a quasicrystalline distribution of defects. A Penrose tiling represents this  substrate
 which act on the thin film. Our Penrose substrate is supposed to be composed
of domains having all the same vertical crystallographic
orientation but different horizontal orientations
due to its rotational defects: it is textured structure. Each vertex of the substrate corresponds
to an elementary quantity of matter (i.e. an elementary domain) of variable height $h$ and crystallographic
orientation $c$, in the thin film.
Due to the quasi crystalline characteristics of the Penrose lattice,
each of these elementary domains has a number of very next
neighbours ranging from  5 to 7. The geometry of these domains corresponds
to the Voronoi tiling of the Penrose lattice.

Beside the action of our substrate on the thin film, stress may  appear because of the surface
morphology of the thin film.
Considering a simple square wave surface morphology, where  the sample
is stressed parallely to the substrate (for example because of the defects of the substrate), the change in energy going from
a flat surface to the square wave surface is roughly \cite{srolo}:
\begin{equation}
\Delta E=\frac{- \sigma^2}{2 A} \frac{h \lambda}{2} +2 \gamma_c
\end{equation}
where $\sigma$ is the stress in the bulk, $\gamma_c$ is the surface energy,
and $A$ is the modulus of the energy of where we have assumed
that the stress in the interior of the square protrusions are zero.
$\lambda$ corresponds to the width of one protrusion and $h$
is the height of the protrusions.

Therefore, we consider here two aspects which contribute to the energy of our thin film
consisting of crystal species: the grain boundary energy (which is here
equivalent to the interfacial energy between two  elementary domains
of different crystallographic orientations) and  the surface energy (corresponding
here to the height of each elementary domain).
For our system of $N$ lattice domains, the energy of this system
can be written as:
\begin{equation}
E=\frac{1}{2} B\sum_{i=1}^N \sum_{j=1}^{NN} (c_i-c_j) +
\frac{1}{2}D \sum_{i=1}^{N} \sum_{j=1}^{NN} (h_i-h_j)+2 \gamma_c
\end{equation}
where the first term of right hand side of the equality corresponds
to the total interfacial energy including grain boundary energy and the second term corresponds
to the total surface energy (see equation (1)). $NN$ is the nearest neighbours of a lattice domain
(varying between 5 and 7). $B$ scales the interfacial energy between two
elementary domains, the numerical values of $c$ range from the lowest interfacial
energy with respect to the substrate, to the highest.
Indeed, the crystallographic orientation $c$ is proportional to the interfacial
energy due to the mismatch of crystallographic orientation between two domains.
$B(c_i-c_j)$ corresponds to the interfacial energy between 
two elementary domains, taking into account the interplay with the substrate
(the value of $c$ with respect to the substrate for which
the crystallographic orientation is constant)
as well as the interplay between two elementary domains (the difference
between two values of $c$).
 $D$ is proportional to the surface energy obtained
for different heights of the elementary domains.
$h_i-h_j$ represents the height difference between two elementary
domains, hence $D (h_i-h_j)$ is the surface energy associated
to the difference in height between two nearest neighbours domains.

For Monte Carlo simulations of single phase films, only one type of event, namely lattice domain reorientation, was considered \cite{srolo2,srolo3}. In our model, the height
of each elementary domain is also submitted to changes.
Thus, a species at domain $i$ may change its orientation with respect of its nearest
neighbour but may also exchange a certain quantity of matter 
represented by a change in the height of the given elementary domain
and its nearest neighbour.
Simulation was performed on a Penrose lattice of 2549 domains.
Periodic boundary conditions were added abruptly on each boundary
of the lattice.
In our model, each domain owns two states $(c,h)$.
Here, $c$ represents the domain orientation with $c=1$ to 4.
By this way, only $c=1$ crystallographic orientation will
be favorable energetically, assuming that this orientation
is that of the substrate and the lowest one.
$h$ has its value enclosed between 1 and 10, assuming that physically,
no elementary domain will have a height larger than 10.

To simulate the islanding of our thin film, prior to simulation,
all elementary domains were assumed to have a height of 1
and a random (enclosed randomly between 1 and 4) crystallographic orientation.
After such initialization, the Monte Carlo algorithm works according
to the following rules. First, a lattice domain is chosen at random
for two independent events (crystallographic reorientation
and height exchange) occurring. Second, for each of those preceding events,
the total change in system energy $\Delta E$ is calculated using equation (2)
and the Metropolis transition function is used to obtain the corresponding
occurring probability:
\begin{eqnarray}
P=1 if  \Delta E <= 0, \\
P=\exp( \frac{- \Delta E}{k_BT}) if \Delta E > 0
\end{eqnarray}
where $k_B$ is the Boltzmann constant and $T$ is the simulation
temperature. Third, for each event, a random number $\xi$ is generated
in the interval 0-1. If $P>\xi$, then the event is made.
Otherwise, the event is rejected and the configuration remains
unchanged. Note that, as height exchange and crystallographic
reorientation are two independent events, it may occur
that a domain changes its height but not its orientation and inversely.
\section{Results and Discussion}
In figure 1, one can see the temporal evolution of a thin film
deposited on a Penrose lattice. The corresponding time is expressed in
Monte Carlo steps MCS (one Monte Carlo step corresponds to the selection
of one domain allowed to change either orientation or height or both of them).
First, there is one small peak appearing (represented by one circle
of larger radius than others): one domain increases its height while its very next neighbors have their height decreased to zero. Then a few dozen peaks appear,
disconnected from each other. Finally, the structure at equilibrium
(i.e. when the structure does not change any more in average)
corresponds to a more or less connected array of islands surrounded by hollow
zones (i.e. zones where the substrate is no more covered 
in figure 1 and in figure 2).

The analysis of the boundary of our sample shows
that there are no more islands there than in the center of it.
Therefore, we can say that taking abrupt periodic boundary conditions
is valid, and that it is not necessary to fit the tiling
on one border with the other border in order that the boundary
conditions maintain an exact Penrose tiling.
This is logical, as the number of nearest neighbours on the borders
are also ranging from 5 to 7 in average with our boundary conditions.

In figure 3, we present the evolution of the the number of domains
having a given height ($h=1,2,3,4$ or 5) as a function of time
expressed in MCS per domain and averaged over 10 different realizations.

In figure 4 one can see the evolution of the number of elementary
domains having a given crystallographic orientation,
as a function of time expressed in MCS.
It is clear that the number of domains with an orientation
corresponding to the lowest interfacial energy with respect
to the substrate, increase while the others decrease.
This transformation  is very fast: as one can see in figure 4,
almost all the elementary domains have an orientation $c=1$
after 100 MCS per domain.

The effect of the substrate is clear as the number of islands
located on vertex with 7 neighbours is  larger than the number
of islands located on vertex with 5 or 6 neighbours (see for that
figure 5). One can see also in figure 5 that the domains with 5 or 6
neighbours have a larger probability to have their
height equal to zero.
The mechanism of growth of islands on the heptagonal domains
is easy to understand: as the number of neighbours is larger
for these domains, the probability for them to grow in height
is larger because exchanges of matter with neighbours 
 are more numerous.

Let us discuss the effect of the substrate in the light of
its quasi crystalline distribution of domains and of geometrical frustration
\cite{mosseri}.
One can consider an assembly of random tilings, associated to a space
of discrete configuration, with a minimal distance which
corresponds, in the structure of the substrate to simple topological rearrangements (called flips) \cite{elser}.
These tilings are derived from canonical tilings, obtained by projection
of an hypercubic array in a higher dimension on the plane.
Here, our thin film has the same domain distribution as the substrate
and can neither change it. As the domains in the thin film
can not change their number of neighbours, i.e. their form,
the simplest way to reduce geometrical frustration where it is high (on heptagons)
is to grow in height.

Finally, the Penrose substrate leads to an intrinsic instability.
This can be proved by the theorem of Herring \cite{herring}:
if a given macroscopic surface of a crystal does not coincide
in orientation with some portion of the boundary
of the equilibrium shape, there will exist a hill and valley
structure which has a lower free energy than a flat surface, while if
the given surface does occur in the equilibrium shape, no hill and valley
structure can be more stable.
Here, it is evident, that our initial plane film surface is not
stable do to quasi-crystalline characteristics of the underneath
substrate. Hence, the most stable structure of the film
is a 'hill and valley ' structure, i.e. several islands
located on the substrate.
This instability comes from two characteristics of our thin film:
the initial polycrystalline structure of the film and
the rotational defects of the substrate.
\section{Conclusion}
We computed the thermal relaxation of a polycrystalline
thin film deposited on a Penrose tiling substrate.
An islanding process occurs during this thermal treatment,
because of the instability of the crystalline structure
of the film as well as of the structure of the substrate
itself.
The islands are located preferentially on the rotational
defects corresponding to heptagons where the geometrical frustration is larger.
As a conclusion, we can say that controlling
the number of rotational defects on a substrate could lead 
to controlling an islanding process.
\begin{figure}
\caption{Representation of the thin film evolution for $k_BT=1$
and for 4 different Monte Carlo steps  (a) $t=1$MCS, (b)$t=1500$MCS, (c) $t=10^5$ MCS and (d) $t=10^6$MCS. The height of each domain is proprotional
to the radius of the circle located on a vertex of the Penrose lattice}
  \end{figure}
  
\begin{figure}
\caption{Representation of the thin film for $k_BT=1$
and for $t=10^6$MCS. An interpolation has been made between
the domains in order not to have abrupt limits
between each island. (a) top view with grey scale corresponding to heights
(b) perspective view}
\end{figure}

\begin{figure}
\caption{Number of domains for heights $h=1,2,3,4,5$ and 6 as
a function of time expressed in MonteCarlo steps 
}
\end{figure}

\begin{figure}
\caption{Number of domains for crystallographic orientation $c=1,2,3$ and 4
as
a function of time expressed in MonteCarlo steps
}
\end{figure}

\begin{figure}
\caption{Number of domains centered on heptagons, hexagons and pentagons
as a function of the height of the corresponding elementary domains}
\end{figure}
\pagebreak

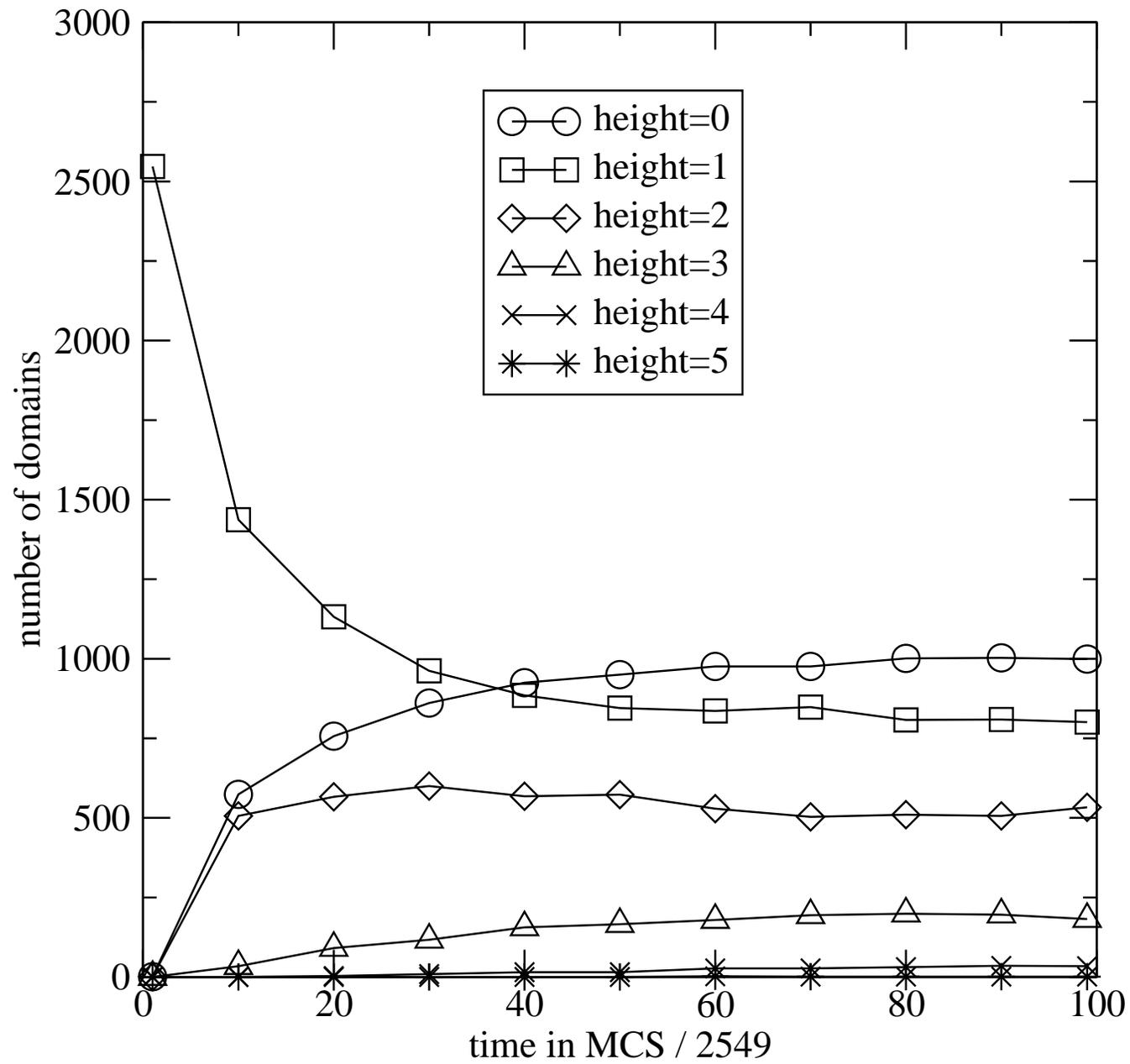
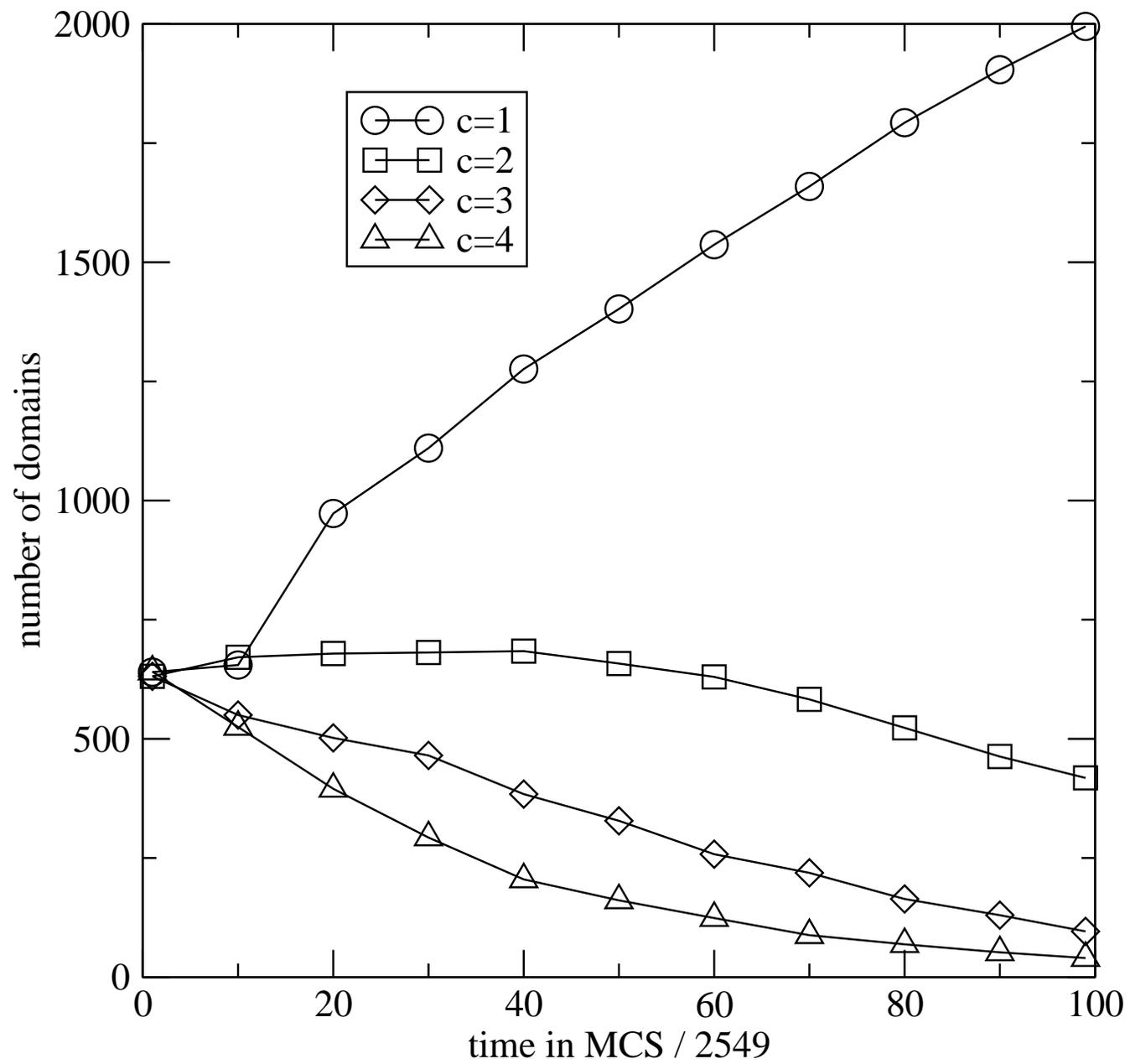
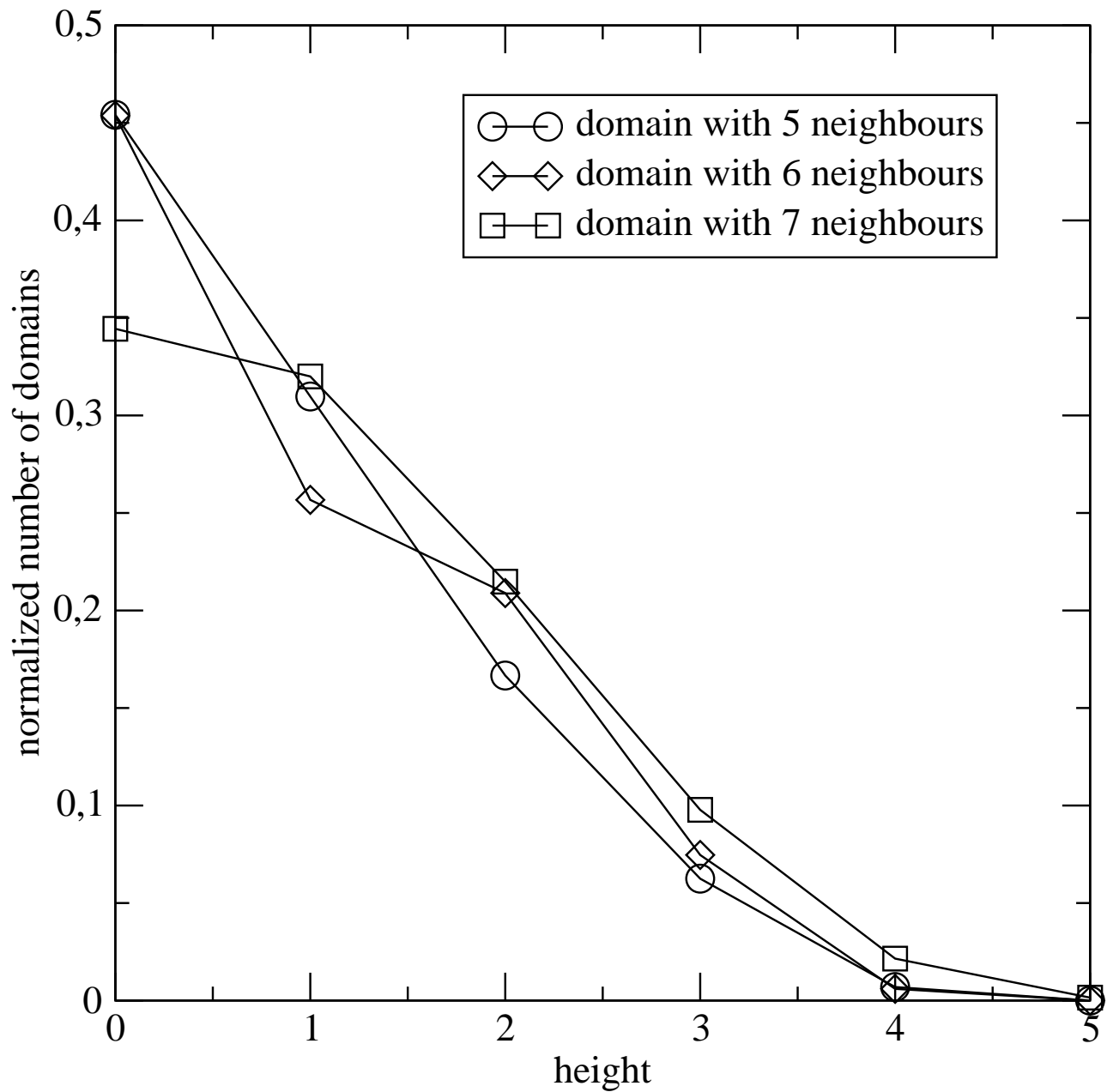
\end{document}